\documentclass{article}
\usepackage{ijcai21}

\usepackage{times}

\usepackage{soul}
\usepackage{url}
\usepackage[hidelinks]{hyperref}
\usepackage[utf8]{inputenc}
\usepackage[small]{caption}
\usepackage{graphicx}
\usepackage{amsmath}
\usepackage{booktabs}
\usepackage{todonotes}
\title{A Survey on Predicting the Factuality and the Bias of News Media}

\author{%
Preslav Nakov$^1$\footnote{Contact Author}\and
Husrev Taha Sencar$^1$\and
Jisun An$^2$\and
Haewoon Kwak$^2$
\affiliations
$^1$Qatar Computing Research Institute, HBKU\\
$^2$Singapore Management University\\
\emails
\{pnakov,hsencar\}@hbku.edu.qa,
\{jisunan,hkwak\}@smu.edu.sg
}
\begin{document}

\maketitle

\begin{abstract}
The present level of proliferation of fake, biased, and propagandistic content online has made it impossible to fact-check every single suspicious claim or article, either manually or automatically. Thus, many researchers are shifting their attention to higher granularity, aiming to profile entire news outlets, which makes it possible to detect likely ``fake news'' the moment it is published, by simply checking the reliability of its source. Source factuality is also an important element of systems for automatic fact-checking and ``fake news'' detection, as they need to assess the reliability of the evidence they retrieve online.
Political bias detection, which in the Western political landscape is about predicting left-center-right bias, is an equally important topic, which has experienced a similar shift towards profiling entire news outlets. Moreover, there is a clear connection between the two, as highly  biased media are less likely to be factual; yet, the two problems have been addressed separately.
In this survey, we review the state of the art on media profiling for factuality and bias, arguing for the need to model them jointly. We further discuss interesting recent advances in using different information sources and modalities, which go beyond the text of the articles the target news outlet has published. Finally, we discuss current challenges and outline future research directions.
\end{abstract}

\section{Introduction}

%\todo[inline]{We need to rephrase the intro a bit; the text below is too similar to what we had in a previous paper.}

The rise of the Web has made it possible for anybody to create a website or a blog and to become a \textit{news medium}.
This was a hugely positive development as it elevated freedom of expression to a whole new level, allowing anybody to have their voice heard.
With the subsequent rise of social media, anybody could potentially reach out to a vast audience, something that until recently was only possible for major news outlets.  
One of the consequences was a \textit{trust crisis}: with traditional news media stripped off their gate-keeping role, the society was left unprotected against potential manipulation.

The issue became a general concern in 2016, a year marked by micro-targeted online disinformation and misinformation at an unprecedented scale, primarily in connection to Brexit and the US Presidential election.
These developments gave rise to the term ``fake news.''

%In an attempt to solve the trust problem,
Several initiatives, such as PolitiFact, Snopes, FactCheck, and Full Fact, have been launched to fact-check suspicious claims manually.
However, given the scale of the proliferation of false information online, it became clear that it was unfeasible to fact-check every single suspicious claim, even when this was done automatically, not only for computational reasons but also due to timing.
In order to fact-check a claim manually or automatically, we need to verify the stance of mainstream media concerning that claim and/or the reaction of users on social media.
Accumulating this evidence takes time, and delay means more potential sharing of the malicious content.
A study has shown that for some very viral claims, more than 50\% of the sharing happens within the first ten minutes after posting the micro-post on social media~\cite{zaman2014}, and thus timing is of utmost importance.  Moreover, an extensive recent study has found that ``fake news'' spreads six times faster and reaches much farther than real news~\cite{Vosoughi1146}.

A much more promising alternative is to focus on the source and profile the medium that initially published the news article.
The idea is that media that have published fake or biased content in the past are more likely to do so in the future.
Thus, profiling media in advance makes it possible to detect likely ``fake news'' the moment it is published by simply checking the reliability of its source. 

Estimating the reliability of a news medium source is important for tasks such as fact-checking a claim \cite{CIKM2020:FANG},
%,DBLP:conf/aaai/NguyenKLW18}, 
and it also gives an important prior when solving article-level tasks such as ``fake news'' and click-bait detection.
%\cite{Hardalov2016,RANLP2017:clickbait,desarkar-yang-mukherjee:2018:C18-1,prezrosas-EtAl:2018:C18-1}. %brill2001online,finberg2002digital,Pan:KG:2018

There have been several surveys on ``fake news'' \cite{Shu:2017:FND:3137597.3137600,10.1145/3395046,CardosoDurierdaSilva2019}, misinformation \cite{Islam2020}, fact-checking \cite{thorne-vlachos:2018:C18-1,Kotonya2020}, truth discovery \cite{Li:2016:STD:2897350.2897352}, and propaganda detection \cite{da2020survey}. However, they have focused either on individual claims or on articles; in contrast, here we survey research on profiling entire news outlets for factuality and for bias.

\section{Factuality}
\label{sec:factuality}

% Journalists, online users, and researchers are well-aware of the proliferation of false information, and thus topics such as credibility and fact-checking are becoming increasingly important.
% For example, the ACM Transactions on Information Systems journal dedicated, in 2016, a special issue on Trust and Veracity of Information in Social Media \cite{Papadopoulos:2016:OSI}.

% There have also been some related shared tasks such as the SemEval-2017 task~8 on Rumor Detection \cite{derczynski-EtAl:2017:SemEval}, the CLEF-2018 lab on Automatic Identification and Verification of Claims in Political Debates \cite{clef2018checkthat:overall}, and the FEVER-2018 task on Fact Extraction and VERification \cite{thorne-EtAl:2018:N18-1}.

Veracity of information has been studied at different levels:
(\emph{i})~claim-level (e.g.,~\emph{fact-checking}),
(\emph{ii})~article-level (e.g.,~\emph{``fake news'' detection}),
(\emph{iii})~user-level (e.g.,~\emph{hunting for trolls}), and
(\emph{iv})~medium-level (e.g.,~\emph{source reliability estimation}).
Our primary interest here is in the latter.

At the claim-level, fact-checking and rumor detection have been primarily addressed using information extracted from social media, i.e.,~based on how users comment on the target claim \cite{Castillo:2011:ICT:1963405.1963500,kochkina-liakata-zubiaga:2018:C18-1}.
%\cite{Canini:2011,Castillo:2011:ICT:1963405.1963500,Ma:2015:DRU,ma2016detecting,PlosONE:2016,P17-1066,dungs-EtAl:2018:C18-1,kochkina-liakata-zubiaga:2018:C18-1}.
A set of web pages and snippets from search engines have also been used as a source of information \cite{RANLP2017:factchecking:external}.
%\cite{mukherjee2015leveraging,popat2016credibility,Popat:2017:TLE:3041021.3055133,RANLP2017:factchecking:external,AAAI2018:factchecking,baly-EtAl:2018:N18-2}.
In either case, the most important information for the claim-level tasks are
\emph{stance} (does a tweet or a news article agree or disagree with the claim?) and
\emph{source reliability} (do we trust the user who posted the tweet or the medium that published the news article?).
%Other important features are linguistic expression, meta information, and temporal dynamics.

% Stance detection has been addressed as a task in its own right, where models have been developed based on data from
% the Fake News Challenge \cite{riedel2017simple,thorne-EtAl:2017:NLPmJ,NAACL2018:stance,hanselowski-EtAl:2018:C18-1}, or from
% SemEval-2017 Task~8 \cite{derczynski-EtAl:2017:SemEval,dungs-EtAl:2018:C18-1,ZubiagaKLPLBCA18}.
% It has also been studied for other languages such as Arabic \cite{DarwishMZ17,baly-EtAl:2018:N18-2}.

%Unlike stance detection, 
The problem of source reliability remains largely under-explored. 
In the case of social media and community fora, it concerns modeling the user. In particular, there has been research on finding opinion manipulation \emph{trolls}, paid \cite{Mihaylov2015ExposingPO} or just perceived \cite{Mihaylov2015FindingOM},
%,mihaylov-nakov:2016:P16-2,AAAI2018:factchecking}, %InternetResearchJournal:2018,
\emph{sockpuppets} \cite{Maity:2017:DSS:3022198.3026360}, \emph{Internet water army} \cite{Chen:2013:BIW:2492517.2492637}, and \emph{seminar users} \cite{SeminarUsers2017}.
In the case of the Web, it is about the trustworthiness of the source (the URL domain, the medium).
The latter is our focus here.

In early work, the source reliability of news media has often been estimated automatically based on the general stance of the target medium with respect to known true/false claims, without access to gold labels about the overall medium-level factuality of reporting \cite{mukherjee2015leveraging}.
%,popat2016credibility,Popat:2017:TLE:3041021.3055133,Popat:2018:CCL:3184558.3186967}.
%The assumption is that reliable media agree with true claims and disagree with false ones, while for unreliable media it is mostly the other way around.
%The trustworthiness of Web sources has also been studied from a Data Analytics perspective. For instance, \cite{Dong:2015:KTE:2777598.2777603} proposed that a trustworthy source is one that contains very few false claims.

More recent work has addressed the task as one on its own right. \cite{baly2018predicting} used gold labels from Media Bias/Fact Check,\footnote{\url{http://mediabiasfactcheck.com}} and a variety of information sources: articles published by the medium, what is said about it on Wikipedia, metadata from its Twitter profile, URL structure, and traffic information. In follow-up work, \cite{source:multitask:NAACL:2019} used the same representation to jointly predict media factuality and bias on an ordinal scale, using a multi-task ordinal regression setup.
Finally, \cite{baly-etal-2020-written} extended the information sources to include Facebook followers and speech signals from the news medium's channel on YouTube (if any). 
Finally, \cite{hounsel2020identifying} proposed to use domain, certificate, and hosting information of the website infrastructure.

\section{Bias}
\label{sec:bias}

\subsection{A Variety of Dimensions in Media Bias}
Compared to factuality, which is decided by whether a piece of information is true or not, media bias has more complex dimensions. For the last few decades, many scholars have conceptualized media bias in different ways. 
For instance, a bias can be defined as ``imbalance or inequality of coverage rather than as a departure from truth''~\cite{stevenson1973untwisting}. % in comparison of presidential election news coverage
They particularly note that a departure from truth, as a bias, can be measured only when the accurate record of the event is available (e.g., trial transcript). 

A different definition, ``Any systematic slant favoring one candidate or ideology over another'' \cite{waldman1998newspaper}, is proposed to capture various other dimensions rather than coverage imbalance, such as favorability conveyed in visual representations (i.e., news photos).  For example, smiling, speaking at the podium, cheering crowd, and eye-level shots are preferred over frowning, sitting, being alone, and shots from above, respectively. 

D'Alessio and Allen reviewed 59 quantitative studies about partisan media bias in presidential elections~\cite{d2000media}, and based on this analysis, they proposed to categorize media bias into the following three types: (\emph{i})~\emph{gatekeeping bias}, where editors and journalists `select' the stories to report, (\emph{ii})~\emph{coverage bias}, where the amount of news coverage (e.g., the length of newspapers articles, or the time given on television) each party receives is systematically biased to one party at the expense of the other one, and (\emph{iii})~\emph{statement bias}, where news media interject their attitudes or opinions in the news reporting. 
Groeling proposed a more relaxed concept of media bias, which is ``a portrayal of reality that is significantly and systematically (not randomly) distorted,'' to take a variety of media bias dimensions into account~\cite{groeling2013media}. In particular, he focused on two main forms of media bias\textemdash \emph{selection bias} (i.e., what to cover) and \emph{presentation bias} (i.e., how to cover it)\textemdash driven by the choices of newsmakers.  

\paragraph{Selection bias} or gatekeeping bias, has been studied in various ways, including qualitative interviews or surveys of journalists and editors about the decision making process they use to select the news stories in their newsroom~\cite{tandoc2014journalism}. 
Data-driven research on selection bias follows the common steps: (\emph{i})~collect news articles (for newspapers or online news) or transcripts (for television news) for a target period, (\emph{ii})~conduct content analysis to find the news coverage of politicians, parties, events, etc. Sometimes the tone of the news articles can be studied (i.e., negative news stories are more frequently reported or selected by the editors compared to positive news)~\cite{soroka2012gatekeeping}, and (\emph{iii})~identify systematic biases by comparing their news coverage. An exhaustive database of news stories is, thus, essential for selection bias research. While commercial databases, such as Lexis Nexis, have been widely used~\cite{soroka2012gatekeeping}, %,padgett2019seen %gilens2000corporate,boykoff2004balance,
publicly available datasets, such as GDELT or Google News, start to get attention~\cite{kwak2014first,boudemagh2017news,kwak2018we} and are getting validated by comparing multiple sources~\cite{weaver2008finding,kwak2016two}. 

\paragraph{Presentation bias} has been characterized from diverse perspectives, including framing~\cite{entman2007framing}, visuals~\cite{barrett2005bias}, sources~\cite{baum2008new}, tone~\cite{soroka2012gatekeeping}, and more.

\subsection{Framing Bias}

\emph{Framing} refers to a process that highlights a certain aspect of an event or an issue more than the others~\cite{entman1993framing}. 
Emphasizing an issue's particular aspect can deliver a distorted view toward the issue even without the use of biased expressions.

Framing biases have been typically studied at issue level. 
Researchers collect news articles about a particular issue or event, conduct manual content analysis on them, and build a frame detection model~\cite{baumer2015testing}. %morstatter2018identifying
Although this approach successfully characterizes diverse frames, it is not trivial to compare media's framing across different issues. 

The Media Frames Corpus (MFC) was proposed to address this limitation. It contains articles annotated with 15 generic frames (including \emph{others}) across three policy issues~\cite{card-etal-2015-media}. 
Several studies have demonstrated reasonable prediction performance of the general media frames with different datasets~\cite{field-etal-2018-framing,kwak2020systematic}. 
These 15 general frames were also used for identifying frames in political discourse on social media~\cite{johnson-etal-2017-leveraging}. 
General media frames are often customized to a specific issue by adding issue-specific frames~\cite{liu-etal-2019-detecting}, even though doing so somewhat contradicts the original motivation of using general media frames, namely to be able to compare frames across various issues.

\subsection{News Slant}

As a related concept to framing, news \emph{slant} was proposed to characterize how the framing in news reports favors one side over the other~\cite{entman2007framing}. The media-level slant thus could be different across issues~\cite{ganguly2020empirical}.

A variety of methods have been proposed to quantify the extent of news slant in traditional news media by (\emph{i})~linking media outlets to politicians with known political positions, (\emph{ii})~directly analyzing news content, and ~(\emph{iii})~using shared audience among media outlets.
For example, Groseclose and Milyo assigned an ADA (Americans for Democratic Action) score for each media outlet by investigating co-citations of think-tanks by members of Congress and media outlets~\cite{groseclose2005measure}. 
Genzkow and Shapiro proposed an ideological slant index of news media in a seminal study~\cite{gentzkow2010what}. The news slant is measured by the extent of phrases in news coverage that are more frequently used by one political party (i.e., Democratic or Republican) congress members than by another one in the 2005 Congress Record. Their frequency-based approach successfully finds politically charged phrases such as \emph{death tax} or \emph{war on terror} Republicans and associated media and \emph{estate tax} or \emph{war in Iraq} by Democrats and associated media, and they further computed media a slant index for 433 newspapers. The choice of words by political party members and news media can be considered framing as well because they purposely highlight some aspect of the issue over other ones. 
An et al. proposed a method to compute media slant scores by measuring distances between media sources by their mutual followers on Twitter and mapping them to a two dimensional space~\cite{an2011media,an2012visualizing}. 
\cite{stefanov-etal-2020-predicting} identified the political leanings of media outlets and influential people on Twitter based on their stance on controversial topics. They built clusters of users around core vocal ones based on their behaviour on Twitter such as retweeting, using a procedure proposed in \cite{ICWSM2020:Unsupervised:Stance:Twitter}.

%Although media bias has been studied from diverse perspectives, these days it typically refers to an ideological slant toward the Democratic or the Republican party due to its simplicity and importance of media bias in political contexts.  

%\todo[inline]{Cite here EMNLP-2018, multitask ordinal regression NAACL-2019, Youtube work, ACL-2020}

\begin{table}[h!]
\footnotesize
\centering
\begin{tabular}{ll}
\toprule
\textbf{Bias Type} & \textbf{Sample Cues} \\
\midrule
Factives & realize, know, discover, learn\\ %notice, 
Implicatives & cause, manage, hesitate, neglect \\ %allow, 
Assertives & think, believe, imagine, guarantee\\ %, seem
Hedges & approximately, estimate, essentially\\ %likely, 
Report-verbs & argue, admit, confirm, express\\ %acknowledge, 
Wiki-bias & capture, create, demand, follow\\ %, hold
\midrule
Modals & can, must, will, shall\\
Negations & neither, without, against, never, none\\
\midrule
Strong-subj & admire, afraid, agreeably, apologist\\
Weak-subj & abandon, adaptive, champ, consume\\ %big, 
Positives & accurate, achievements, affirm\\ %, amiable
Negatives & abnormal, bankrupt, cheat, conflicts\\
\bottomrule
\end{tabular}
\caption{Some cues for various bias types.}
\label{table:bias_types}
\end{table}

\section{Basis of Prediction}

\subsection{Textual Content}

\subsubsection{Representation}

The most natural representation for a source is as a sample of articles it has published, which in turn can be represented using linguistic features or as continuous representations.

\textit{Linguistic Features:}
These features focus on language use, and they have been shown to be useful for detecting fake articles, as well as for predicting the political bias and the factuality of reporting of news media~\cite{DBLP:journals/corr/abs-1803-10124,baly2018predicting}. For example,  \cite{DBLP:journals/corr/HorneA17} showed that ``fake news'' pack a lot of information in the title (as many people do not read beyond the title, e.g., in social media), and use shorter, simpler, and repetitive content in the body (as writing fake information takes a lot of effort).
Such features can be generated based on the Linguistic Inquiry and Word Count (LIWC) lexicon and used to distinguish articles from trusted sources vs. hoaxes vs. satire vs. propaganda \cite{pennebaker2001linguistic}. They can be also modeled using linguistic markers \cite{AAAI2018:factchecking} such as \textit{factives} from~\cite{hooper1974assertive}, \textit{assertives} from~\cite{hooper1974assertive}, \textit{implicatives} from~\cite{karttunen1971Implicatives},  \textit{hedges} from~\cite{hyland2005metadiscourse}, \textit{Wiki-bias} terms from~\cite{recasens2013linguistic}, \textit{subjectivity} cues from~\cite{Riloff:2003:LEP:1119355.1119369}, and 
\textit{sentiment} cues from~\cite{Liu:2005:OOA:1060745.1060797}; see Table~\ref{table:bias_types} for examples.
There are 141 such features implemented in the NELA toolkit~\cite{DBLP:journals/corr/abs-1803-10124}, grouped in the following categories:

\begin{itemize}
\item \textbf{Style}: part-of-speech tags, use of specific words (function words, pronouns, etc.), and features for clickbait title classification;
%from \cite{clickbait:2016};
\item \textbf{Complexity}: type-token ratio, readability, number of cognitive process words (identifying discrepancy, insight, certainty, etc.);
\item \textbf{Bias}: features modeling bias using lexicons
%~\cite{recasens2013linguistic,mukherjee2015leveraging}
and subjectivity as calculated using pre-trained classifiers;
%~\cite{horne2017identifying};
\item \textbf{Affect}: sentiment scores from lexicons
%\cite{recasens2013linguistic,mitchell2013geography} 
and full systems;
%\cite{gilbert2014vader};
%\item \textbf{Engagement}: number of shares, reactions, and comments on Facebook;
%\item \textbf{Topic}: lexicon features to differentiate between science topics and personal concerns;
\item \textbf{Morality}: features based on the Moral Foundation Theory \cite{graham2009liberals} and lexicons;
%\cite{lin2017acquiring};
\item \textbf{Event}: features modeling time and location.
\end{itemize}

\textit{Embedding representations:}
An alternative way to represent an article is to use embedding representations, typically based on BERT~\cite{devlin2018bert}. 
This can be done without fine-tuning, e.g., by encoding an article (possibly truncated, e.g., BERT can take up to 512 tokens as an input) and then averaging the word representations extracted from the second-to-last layer.\footnote{This is common practice, since the last layer may be biased towards the pre-training objectives of BERT.}
Alternatively, one can use pre-trained sentence encoders such as Sentence BERT~\cite{reimers2019sentence} or the Universal Sentence Encoder (USE) \cite{cer-etal-2018-universal}.
Finally, one can obtain representations that are relevant to the target task, e.g., by fine-tuning BERT to predict the label (bias or factuality) of the medium that an article comes from, in the form of distant supervision~\cite{baly-etal-2020-written}. One issue with distant supervision is that the model can end up learning to detect the source of the target news article instead of predicting its factuality/bias, which can be fixed using adversarial media adaptation and a specially adapted triplet loss \cite{baly-etal-2020-detect}.

\subsubsection{Aggregation}

In order to obtain a representation/prediction for an entire medium, there is a need to aggregate the representations/predictions for its articles.

\textit{Averaging article-level representations:}
One could average the representations for all articles to obtain a representation for a medium, which can then be used to train a medium-level classifier.
Using arithmetic averaging is a good idea as it captures the general trend of articles in a medium, while limiting the impact of outliers. For instance, if a medium is known to align with left-wing ideology, this should not change if it published a few articles that align with right-wing ideology.

\textit{Aggregating posterior probabilities:}
Alternatively, each article can be represented by a $\mathcal{C}$-dimensional vector that corresponds to its posterior probabilities of belonging to each class $c_i$, $i\in\{1, \dots, \mathcal{C}\}$ of the given task, whether it is predicting the political bias or the factuality of the target news medium.
Finally, these article-level posterior probabilities are averaged in order to aggregate them at the medium level.

% ----------

% Review of different article representation approaches.

% This includes similarity to previously fact-checked claims.

% Joint modeling of metaphor, emotion and political rhetoric~\cite{huguet-cabot-etal-2020-pragmatics}

% using news title, content, and link~\cite{kulkarni-etal-2018-multi}

% article-level ideological slant, which is left, center, or right~\cite{baly-etal-2020-detect}, 

\subsection{Multimedia Content}

We have come to understand events in a far more visual way than we have ever before. 
As a result, multimedia content is now heavily relied upon as a source of news and opinion and has been an important element of almost all news websites.
This dependence, however, also makes multimedia a very effective means for dispensing an intended, and even manipulated, message.
The increasing availability of automated and AI-powered multimedia editing and synthesis tools, combined with massive computational power, makes such capabilities accessible to everyone.

Given that multimedia editors of a news site typically follow a defined workflow when creating, acquiring, editing, and curating content for their pages, this pattern thus adds a crucial dimension to profiling the factuality and the bias of a news source. 
In fact, questions around the origin and the veracity of photographic images and videos have long been the subject of multimedia forensics research \cite{dif}. 

With this objective, several methods have been proposed based on verifying metadata integrity \cite{iuliani2018video,yang2020efficient},
%kee2011digital,
digital integrity
%\cite{korus2017digital,
\cite{cozzolino2018camera},
physical integrity \cite{IulianiFCP17,MaternRS20},
%RiessUNPSA17,PengWDT17,OBrienF12
identification of processing traces \cite{HadwigerBPR19}, and discrimination of synthesized (i.e., GAN generated) media \cite{AgarwalFFA20,verdoliva2020media}.  
%li_etal_wifs18,
Currently, these capabilities have only been sparsely explored in the context of predicting factuality and bias.

Existing work mainly considered characterization of images appearing at trustworthy sources and such obtained from low-factuality news sources. 
These methods have proposed to use visual characteristics of images \cite{jin2016novel}, 
deep-learning visual representations \cite{qi2019exploiting,singhal2019spotfake}, 
%khattar2019mvae,
image provenance information obtained through reverse image search \cite{zlatkova2019fact}, and self-consistency with respect to metadata information \cite{Huh_2018_ECCV}.
Overall, the results of this line of research indicate that multimedia characteristics have a strong potential that has not yet been fully used for news media profiling.

\subsection{Audience Homophily}
%Overview of intelligence derived from different social media platforms.
The well-known homophily principle, ``birds of a feather flock together,'' crucially asserts that similar individuals interact with each other at a higher rate than dissimilar ones.
Therefore, audience representation could be another approach to describe a news media outlet whereby an overall, descriptive characteristic of followers of the outlet is obtained. 
Then, by evaluating the similarity of audience-centric representations with previously categorized news media, the factuality and the bias of the medium in question can be inferred.

\cite{ribeiro2018media} used Facebook's targeted advertising tool to infer the ideological leaning of online media based on the political leaning of the users who interacted with these media, according to Facebook. \cite{an2012visualizing} relied on follow relationships on Twitter to ascertain the ideological leaning of news media and users.
\cite{wong2013quantifying} studied retweet behavior to infer the ideological leanings of online media sources and popular Twitter accounts. \cite{barbera2015birds} proposed a statistical model based on the follower relationships to media sources and Twitter personalities to estimate their ideological leaning.

\cite{stefanov-etal-2020-predicting} predicted the political leaning of media with respect to a topic by observing the users of which side of the debate on a polarizing topic were sharing content from which media in support of their position in the context of that debate. In particular, they constructed a user-media graph and then used label propagation and graph neural networks to derive representations for media, which they used for classification. They further aggregated the leanings across several polarizing topics to come up with a left-center-right polarization prediction.

Following a similar approach, \cite{baly-etal-2020-written} considered three social media platforms for audience characterization.
On Twitter, they proposed to use self-descriptions in publicly accessible profiles of users following the account of a medium.
For each medium, a representation is obtained by encoding the biographic descriptions of Twitter followers and averaging the resulting textual representations.
The second characterization involves how the audience of the medium's YouTube channel responds to each video in terms of number of comments, views, likes and dislikes.
By averaging these statistics over all videos, a medium-level representation is obtained. 
The last audience representation is obtained using Facebook's advertising platform, which is used to obtain demographic information for the audience interested in each medium. 
This data is used to obtain the audience distribution over the political spectrum.
The distribution is then divided into five categories to label each medium accordingly: very conservative, conservative, moderate, liberal, and very liberal.

\subsection{Infrastructure Characteristics}
Beyond textual, visual, and audience features, news sites also exhibit distinct characteristics that relate to the underlying infrastructure 
and technological components deployed to serve their content online. 
In this regard, the prediction problem is analogous to a well-studied one in the cybersecurity domain where the goal has been to identify infrastructure characteristics of malicious domains~\cite{anderson2007spamscatter,invernizzi2014nazca}.
% that are used for malware distribution \cite{wang2013detecting,invernizzi2014nazca}, phishing \cite{james2013detection,mohammad2012assessment,mohammad2014predicting,purwanto2020phishzip}, online scams \cite{konte2009dynamics,hao2016predator,alrwais2017under}, and spamming \cite{anderson2007spamscatter,hao2009detecting}.
Since establishing the infrastructure of a news medium involves several decisions with respect to technological aspects, it is plausible to expect that news media with varying IT practices and different levels of access to IT human resources will differ in their characteristics. 

So far, only a few works exploited this dimension with a focus on network, web design, and data elements of a news website to essentially discriminate new sites based on factuality and bias.
At the network level, \cite{hounsel2020identifying} considered a comprehensive set of features that relate to a website's domain, certificate, and hosting properties. 
In a classification setting with three classes (i.e., disinformation websites, authentic websites, and sites not related to news or politics), their results showed that features related to a website's domain name, registration, and DNS configuration serve as best predictors for classification.  
Concerning the web design aspect, \cite{castelo2019topic} introduced a web page classifier based on several features that govern the structure and style of a page in addition to three categories of linguistic features. 
Their binary classification results (real or fake news) obtained on several datasets showed that the web-markup features consistently perform well and are complementary to linguistic features.

Lastly, at the data level, \cite{fairbanks2018credibility} examined the source of web pages to identify shared data objects, such as mutually linked sites, scripts, and images, across web sites. 
This information is then used to create a shared data object graph. 
By comparing the content level features with the structural properties of the graph, they found that 
the use of mutually shared objects yields better performance in predicting both factuality and bias of a site with a significant difference in the former task. 
Overall, a major advantage of utilizing infrastructure features comes from their content- and audience-agnostic nature. 
Because of this, they allow making reliable predictions when only limited textual and visual content is available and without an established audience interest in a news medium.

\section{Lessons Learned}

%Important to use multiple information sources, what was written, who read it, what is written about it. Yet, text remains the central topic. Factuality and bias should be modeled together.

Factuality and bias have some commonalities as they exert negative influences on the public by delivering information that is deviated from the truth. 
Not surprisingly, some news media purposedly take a biased position in the political landscape and appeal to partisan audiences. This trend becomes apparent in recent years mainly because the news industry becomes more and more competitive.
Many journalists and editors, however, have concerned about their biases in news selection and reporting and try to be neutral or at least report diverse perspectives of an issue.

As the bias can be conveyed in different means, which are text, photos, and videos, through even a very subtle way, the media bias has complex dimensions. Among them, ideological bias is an important conceptualization due to the importance of media bias in a political context. In the US context, the ideological bias could be broadly defined as conservative, center, and liberal. Then, the (ideological) bias prediction task is formulated as predicting whether a given news story, including both text and visual elements, favors one party over the other. 
Reported results so far show that accurate prediction of this ideological bias of a news medium is a far more easier task than assessing factuality.
This is, in fact, not surprising as evaluation of the factuality ultimately depends on the authenticity and the objectivity of the particular claims stated in a news story, essentially requiring verification from other sources and observations. 

Although more sophisticated analysis of the text style and multimedia characteristics may be expected to improve the achievable accuracy, it is evident that there is a big need to complement the textual and visual 
elements of a news medium with others.
In this regard, recent studies have demonstrated the potential of audience homophily and the medium's infrastructure characteristics in bridging the existing performance gap.
The content-agnostic nature of these characteristics  make them further useful in the early discovery and categorization of news media even in the absence of sufficient content.

\section{Challenges and Future Forecasting}

\subsection{Major Challenges}

\paragraph{Ordinal scales:} While the ideological bias (news slant) is typically modeled as left-center-right, there exists a spectrum  within each bias based on bias intensity. A hyperpartisan (an extreme partisan) bias prediction task has been tested to differentiate far-right from right and far-left from left, but it does not model the political bias using an ordinal scale. Difficulties in labeling the bias (i.e., creating ground-truth datasets) by experts or crowdsourcing is a major hurdle for modeling ideological bias as an ordinal variable.

\paragraph{Joint modeling:} There is a well-known connection between factuality and bias.\footnote{\url{http://www.poynter.org/fact-checking/media-literacy/2021/should-you-trust-media-bias-charts/}}
    For example, hyper-partisanship is often linked to low trustworthiness, e.g., appealing to emotions rather than sticking to the facts, while center media tend to be generally more impartial and trustworthy \cite{source:multitask:NAACL:2019}. Thus, it makes sense to model factuality and bias jointly.

\paragraph{Multimodality:} In news reporting, a news photo typically gets high attention from readers. The fact that readers sometimes can understand news stories  from news photos only---even without reading text---indicates that news text and photos are strongly coupled together and deliver relevant information about news stories to readers. Thus, there should be a benefit from modeling news text and photos together to understand their bias and factuality.
%    \item Audience overlap \todo[inline]{expand or cut}

\paragraph{Evaluation granularity:} The label of a news medium is essentially inferred from a sample of observations. This potentially introduces a measurement bias when a news medium does  not exhibit the same reporting behavior with all news items it publishes. This is especially the case for media that have a particular stance in only certain issues~\cite{ganguly2020empirical}. Thus, reliable estimation of factuality and bias labels require analyzing a relatively large amount of content covering a range of issues.
    
\paragraph{Variability in factuality \& bias ratings:} These ratings are inherently not static and may change over time when a news medium takes corrective action to address issues raised by fact-checkers. 
    In other words, the ground truth needed for building a learning approach varies, triggering the need for re-evaluating the performance of proposed approaches. 
    Therefore, there is also a need to take into account the sensitivity of a learning approach to such small but nevertheless inevitable variations. 
    
%    \item are the datasets good? making decision on a single failed check? What about looking at the intent to harm? -- \todo[inline]{expand or cut}

\paragraph{Dataset size:} The existing datasets for media-level factuality and bias are relatively small in size, typically of a few hundred examples, sometimes a few thousand. These are derived from sites, such as Media Bias/Fact Check and AllSides, where domain experts perform a careful manual analysis based on clear guidelines. 
    
\paragraph{Annotation vs. modeling:} One problem is that human annotators judge the factuality of reporting and the bias of media based on criteria that are not easy to automate or based on information that may not be accessible to automatic systems. For example, if a news outlet is judged to be of mixed factuality based on it having failed just 2-3 fact-checks, for an automatic system to arrive at the same conclusion using the same idea, it would have to select for analysis the exact same articles where the false claims were made.
%    The criteria for judging the factuality of a website include a number of factors, e.g., whether the news medium has failed fact-checks. Many of these criteria are not being modeled as this is hard. For example, if a medium was judged to be of mixed factuality based on 2-3 failed fact-checks, an automatic system would need to have access to the exact articles where these claims were made. 
    
\paragraph{Data availability:} Primarily due to copyright issues, there are only a few publicly available datasets of the full text of news for research purposes. Instead, indexed data (e.g., GDELT dataset\footnote{https://www.gdeltproject.org/}) by mentioned actors, events, locations, sources, or tones are available and have been analyzed in many studies. A set of news headlines collected from news websites or aggregated websites (e.g., AllSides\footnote{https://allsides.com/}) are also shared more actively for research purposes. Considering the importance of social media channels in news dissemination, researchers collect and analyze social media posts of official accounts of news media. As social media posts are relatively more informal than news articles to fit for social media audience~\cite{park2021understanding}, more studies are required for understanding their biases and factuality correctly.

\subsection{Future Forecasting}

Based on the challenges mentioned in the previous subsection, we forecast the emergence of the following research directions:

\paragraph{Support for non-English corpora and different political systems:} Most of the studies we review are conducted based on English corpora. More research on bias and factuality in non-English corpora thus will be expected. Recently, various approaches are proposed to accelerate NLP research for resource-scarce languages, such as multilingual word embeddings. We believe that those efforts help conduct bias and factuality research for non-English corpora. One non-technical issue here is that not all the countries have US-like left-center-right political biases. For example, there might exist a multiparty system in some countries. In that case, understanding relevant political biases should be the first step in media bias research.    

\paragraph{Research on video news:} TV news accounts for significant portions of the news industry. Also, the presence of news media becomes strong in video-driven social media platforms over time. To get high user engagements, news media outlets upload short video clips curated for social media use, particularly on existing social media.
Most previous studies on bias in video news have analyzed their transcripts instead of analyzing video directly. Commercial databases, such as Lexis Nexis, or open-source libraries to create subtitles are used to analyze news transcripts. We expect that more studies on analyzing video contents in an end-to-end manner will be presented to fully understand the bias and factuality of video news.
    
\paragraph{Bringing practical implications:} Since the factuality and bias of news media largely influence the public, it is crucial to implement working systems so that readers can benefit from a rich stream of research. Several stand-alone websites, such as Media Bias/Fact Check, AllSides, and Tanbih \cite{EMNLP2019:tanbih},\footnote{https://www.tanbih.org/} aim to make media bias and factuality transparent to end-users, thus promoting media literacy. We expect new tools and services that support more media and languages in the coming years.

\section{Conclusion}

We reviewed the state of the art on media profiling for factuality and bias, arguing for the need to model them jointly. We further discussed interesting recent advances in exploiting different information sources and different modalities, which go beyond the text of the articles the target news outlet has published. Finally, we discussed current challenges and outlined promising research directions.

%\todo[inline]{We are too long. We need to put the references in compressed format, with no address, no long name of conference, etc. Like here: https://www.ijcai.org/Proceedings/2020/0672.pdf}

\clearpage % easier to see how much we need to cut from references

{\small
\bibliographystyle{named}
\bibliography{ijcai21}
}

\end{document}